# The challenge of measuring and mapping the missing baryons

Simon Driver

**The missing baryon problem may now be resolved, but the exact location and physical properties of the diffuse component remains unclear. This problem could be tractable, but requires the combination of new galaxy redshift surveys with new X-ray and radio facilities.**

Big Bang Nucleosynthesis *(BBN)* calculations[1] infer a baryon density today of $\rho_b$ = (4.19 +/- 0.08] x $10^{-28}$ kg/$m^3$ in S.I. units, or $\Omega_b h^2$ = 0.02190 to 0.02271 when expressed as a fraction of the critical density and where $h$ is given by the Hubble Constant ($H_0$) divided by 100 km $s^{-1}$Mpc$^{-1}$. The term baryon represents normal matter formed from odd quark/antiquark combinations, overwhelmingly protons and neutrons. Hence baryons make up the nucleons of the periodic table and all that that entails, including us, the Earth and the stars.

For many years the BBN prediction has presented a major challenge[2] since when we sum the mass of objects comprised of nucleons, that is, the plasma, gas, stars, dust, black holes, we fall short by about 30%. This missing baryon problem[2] is distinct from the dark matter problem, in which five times more mass than implied by BBN is needed to explain cluster velocity dispersions, galaxy rotation curves and the growth of structure, for which a baryon solution is all but ruled out, and a 'beyond the standard model' solution implied.

The BBN estimate of the baryon density sits snugly with the many cosmological-scale constraints, in which dark energy, dark matter, normal matter, the Hubble constant, and other key parameters are simultaneously constrained and gives $\Omega_b h^2$ = 0.02234 to 0.02242[3]. It is the concordance between these independent measurements that is compelling, and which now includes the cosmic microwave background (CMB), type Ia Supernova, baryonic acoustic oscillations, redshift space distortions, weak lensing, and rich cluster constraints. Consequently, the finger of suspicion for the baryon discrepancy is firmly pointed towards our empirical census of the nearby Universe.

Our nearby baryon census derives from large spectroscopic or radio surveys of galaxies. We observe, detect and measure what we see, estimate stellar, dust and gas masses, and correct for distances and volumes. Typically, the baryonic mass bound within the galaxy population adds up to only ~8-11% of the BBN value[2] (see Fig. 1): 4-7% stars ~2% in neutral gas (H, He or $H_2$) and ~2% as plasma in the circumgalactic medium. From studies of rich galaxy clusters, we also detect a hot ($10^5$-$10^7$K) plasma in the intracluster medium. This plasma dominates the baryon content of these very dense structures, is bound to the underlying dark matter halo, and provides a further ~4-5% of the *BBN* value[2].

The remaining component (~85%), is proposed[2,4] to reside in various forms, including a warm and hot ionised medium (WHIM) that surrounds galaxy groups, an intrahalo medium (IHM) that pervades galaxy groups and clusters, a loosely bound medium aligned with the fine filamentary structure, and the general inter-galactic medium (IGM) that is unbound to any particular structure. To quantify the masses of these components we must lean heavily upon simulations and theory[2], adding significant uncertainty to our estimates.

Physically, the IGM plasma is a relic component from the epoch of reionisation, in which the neutral gas that formed after matter-radiation decoupling was ionised by the very first stars. While most of this gas is primordial, some small portion represents material heated or processed through stellar nucleosynthetic reactions, and expelled as metal enriched outflows from galaxy hosts driven via supernova, or other mechanical processes.

These outflows are important, as any metal (or dust) enrichment dramatically changes the propensity of the plasma to cool, impacting the temperature distribution and the mass one might infer from the limited observations. At the extreme level, cooling may result in recombination leading to the formation of diffuse low column density neutral hydrogen[5]. Such a component is generally not formed within simulations due to the complexities of modelling a multi-phase medium, but diffuse HI, comparable in mass to that bound to the galaxies, has recently been seen in the Fornax A cluster by MeerKat[6].

Without clear empirical confirmation of the intrahalo medium, WHIM, filamentary and IGM plasma or neutral gas components, other alternative solutions are occasionally raised. These may argue that the local region of the Universe is under-dense; that we are missing large populations of compact, low mass or diffuse galaxies; that an unidentified baryonic component exists within galaxies such as molecular gas, grey dust, mini black holes, or sub-stellar dwarfs; that the BBN value is wrong due to missing nucleosynthetic pathways or additional neutrino species; or that our methods for measuring stellar, neutral, molecular, or plasma masses are systematically biased low.

Recently, two modes of observation have purported to find the missing baryons, and the results seem to confirm that they are indeed in a low-density ionised form as foreseen. The first comes from high resolution spectra of distant quasars[7,8], where three OVII absorption line systems have been identified and combined with simulations to infer the column density of the intervening plasma. The second comes from the advent of Fast Radio Bursts[9] (FRBs), and while the nature of the FRB emission mechanism is unclear, their origin from external galaxies is confirmed. Here the FRB pulse is delayed through interactions of the radiation with the electrons in a wavelength-dependent manner. Hence the wavelength dispersion of the four FRB peaks, combined with their distances, provides a model-free estimate[9] of the intervening mean electron (plasma) density.

The central two pillars of Fig. 1 show the newly revised baryon budget, demonstrating plausible consistency with the BBN value; nevertheless, the associated uncertainties still remain large at +/− 30%[7] and +/−53%[9] of the BBN value. As more absorption lines and more FRBs are identified, these errors will inevitably reduce, and hopefully converge towards the BBN value. However, neither method is likely to allow for a comprehensive three-dimensional (3D) mapping of the plasma sector, nor will they allow for a detailed determination of key plasma characteristics such as its temperature and density. So how then will we ever measure, map and model the diffuse plasma, stellar and HI sectors given the weakness of their respective signals?

A novel technique pioneered in radio astronomy[10], and gaining traction in the X-ray community[11], may help. Very large wide area spectroscopic and/or radio surveys can produce exquisite 3D maps of the nearby cosmos, in which the galaxies, groups, clusters and filaments are readily identifiable (see Fig. 2). This opens the door to 'stacking'of the X-ray, optical, or 21 cm signals, at the 3D locations of clearly defined sub-samples.

By stacking thousands or tens of thousands of targets one amplifies the signal-to-noise by root-$N$ (where $N$ is the number of objects stacked), assuming the inherent systematic uncertainties in the underlying data are small. This root-$N$ gain has now been realised for galaxy spectra, HI data cubes and, most importantly, X-ray imaging[10]. In particular, stacking of ROSAT data, using mass-selected galaxy sub-samples drawn from the Sloan Digital Sky Survey, has provided a clear detection and quantification of the circumgalactic medium component[10].

Unfortunately, stacking of the galaxies only gets you so far – as discussed earlier, the total baryons bound in galaxies is a small fraction (8-11%) of the BBN value (see Fig. 1). The bigger question, and the challenge, is whether we can extend the concept of stacking to larger, lower density environments, and start to stack samples of galaxy groups, filaments, and even voids, to tease out the signal from the more dominant diffuse components. At present our spectroscopic catalogues of groups and filaments are fairly modest. The Galaxy And Mass Assembly (GAMA) team provide one of the most comprehensive catalogues[12,13], containing over 6,000 galaxy groups[12] with halo masses ranging from $10^{12}M_o$ to $10^{15}M_o$, where $M_o$, is one solar mass, and 2,000 filaments or tendrils[13].

These current group and filament catalogues may be sufficient to demonstrate viability, but will probably be insufficient to fully map the entirety of the diffuse plasma, HI and stellar domains after subdividing by halo mass, redshift, or other parameters. Moreover, the ultimate goal is not just to detect the plasma, stripped stars, and gas, but to characterise them by robustly extracting properties such as the mean density profiles and, for the dominant plasma component, the temperature, requiring high signal-to-noise stacks of multiple spectral lines.

One further advantage, of constructing a baryon inventory in terms of the structures to which they are bound, comes from comparisons to the next-generation hydrodynamical simulations[14]. Here we can `observe' the simulations, by using the same group and filament finding codes and stacking methods, to manage any inherent bias in the methodology. Hence, even if the stacking of voids and filaments remains intractable, the ability to compare observations to simulations as a function of galaxy and group halo mass should be valuable and informative.

Over the next decade, our detailed maps of galaxy groups, filaments and voids is set to grow significantly (see Fig. 2) with the commencement of the Dark Energy Spectroscopic Instrument Bright Galaxy Survey[15] (DESI-BGS) providing a GAMA-depth survey over a sixty times larger area, and ESO's 4MOST Wide Area VISTA Extragalactic Survey[16] (WAVE*S*). WAVES consists of two sub-surveys that will push to lower halo masses over five times the GAMA area (WAVES-Wide), as well as out to a lookback time of eight billion years within a 50 square degree region (WAVES-Deep).

This opportunity to build structural catalogues as priors for the stacking of X-ray, optical/near-infrared and radio data is especially exciting. Not only does it provide a possible pathway to measuring and mapping the plasma, stars, and gas in diffuse regions locally, but over all time as our structural catalogues extend deeper and a new generation of higher-resolution and more sensitive X-ray, optical, infrared, and radio facilities come online. Critical to this process is the need to align where new facilities conduct their deep survey fields, and this alignment requires coherent pan-facility coordination, leadership and vision to maximise the collective science return.

*Simon Driver is at the International Centre for Radio Astronomy (ICRAR) and the International Space Centre (ISC) of the University of Western Australia, Perth, Australia. e-mail: simon.driver@uwa.edu.au*

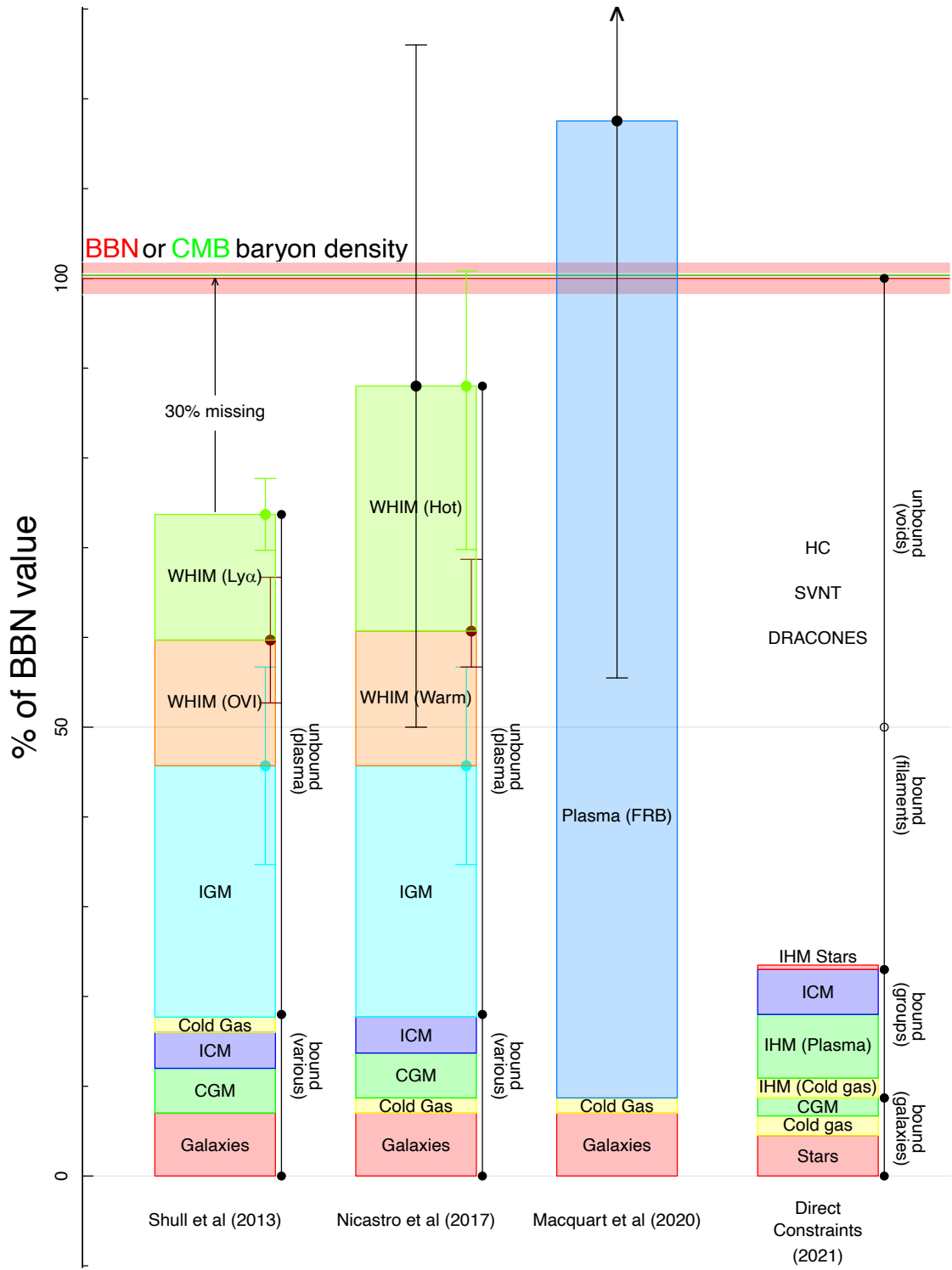

Figure 1 | **Recent estimates of the local baryon census.** Left: Shull et al.[2] first clearly articulated the 30% missing baryon problem. Middle: This problem may have been solved by the recent analyses of Nicastro et al.[7] and Macquart et al.[9], however the uncertainties at this stage are large (and note the upper error bar of Macquart et al. extends significantly beyond the plotting region). Right: A compendium based on the recent literature but now highlighting the structures to which the baryons are bound. The horizontal lines show the current BBN (red) and CMB (green) baryon density constraints: ICM, intracluster medium; CGM: circumgalactic medium; IHM, intrahalo medium.

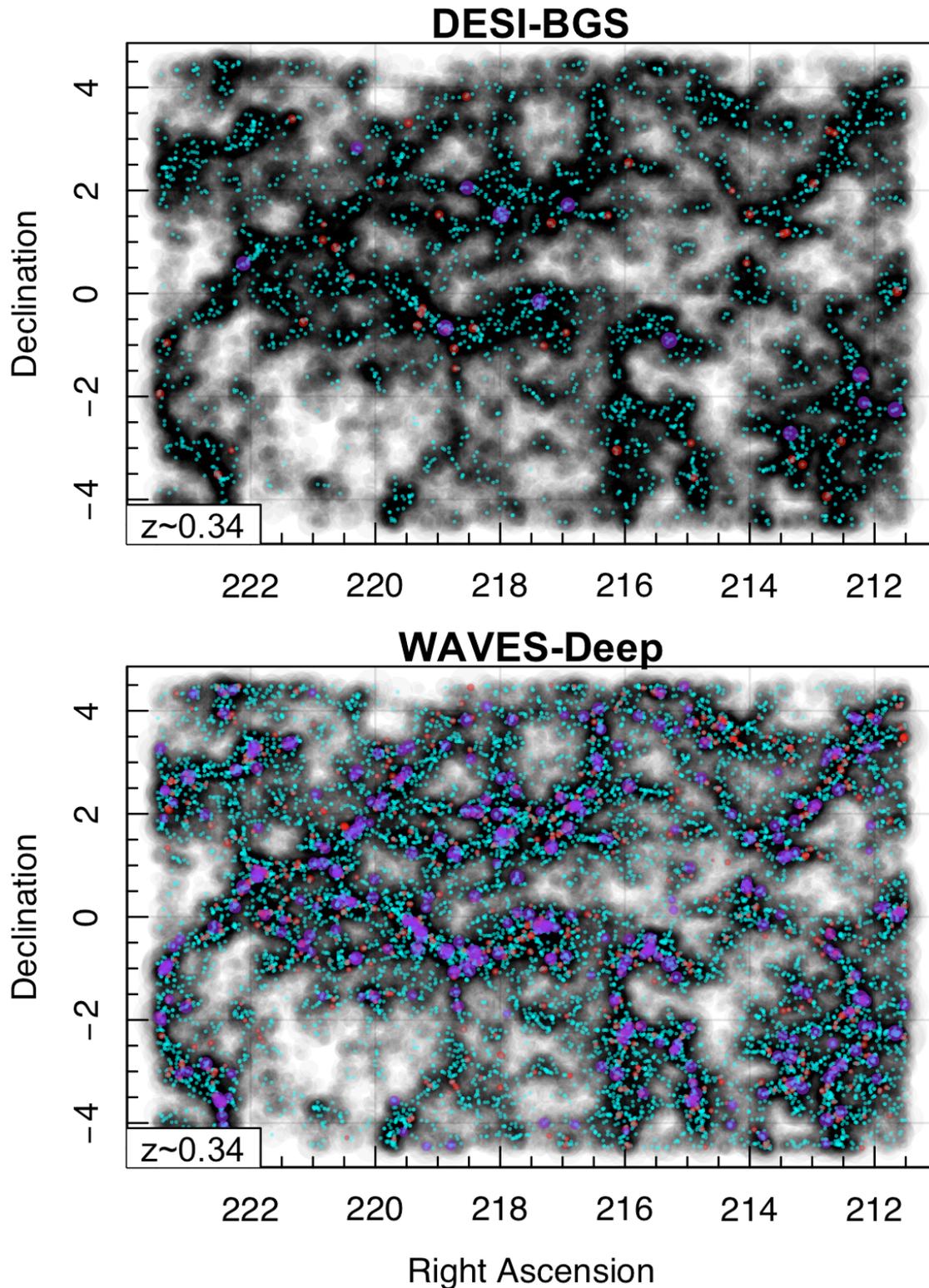

Figure 2 | **The underlying structures that surveys such as DESI-BGS and WAVES will reveal.** The next generation redshift surveys, now underway, will provide unprecedented detail of the underlying dark matter structure (black shading), traceable by the galaxy (cyan dots), pair (red circles), group (purple circles), filament, and void distributions, shown here at redshift ~0.34. The very weak plasma and neutral gas emission associated with these structures can be teased out through the stacking of X-ray and radio spectral lines at the a priori space-time locations of these structures. Based on data from the Shark simulations provided by C. Lagos and C. Power.